\documentclass[reprint,amssymb,amsmath,aip,rsi]{revtex4-1}
\usepackage{bm}
\usepackage{graphicx}
\usepackage[colorlinks=true,linkcolor=blue]{hyperref}
\begin{document}
\title{Higher magnetic-field generation by a mass-loaded single-turn coil}
\author{M. Gen}
\email{gen@issp.u-tokyo.ac.jp}
\affiliation{Institute for Solid State Physics, University of Tokyo, Kashiwa, Chiba 277-8581, Japan}
\author{A. Ikeda}
\affiliation{Institute for Solid State Physics, University of Tokyo, Kashiwa, Chiba 277-8581, Japan}
\author{S. Kawachi}
\affiliation{Tokyo Institute of Technology, Yokohama, Kanagawa 226-8503, Japan}
\author{T. Shitaokoshi}
\affiliation{Institute for Solid State Physics, University of Tokyo, Kashiwa, Chiba 277-8581, Japan}
\author{Y. H. Matsuda}
\affiliation{Institute for Solid State Physics, University of Tokyo, Kashiwa, Chiba 277-8581, Japan}
\author{Y. Kohama}
\affiliation{Institute for Solid State Physics, University of Tokyo, Kashiwa, Chiba 277-8581, Japan}
\author{T. Nomura}
\affiliation{Institute for Solid State Physics, University of Tokyo, Kashiwa, Chiba 277-8581, Japan}
\date{\today}

\begin{abstract}
Single-turn coil (STC) technique is a convenient way to generate ultrahigh magnetic fields of more than 100~T.
During the field generation, the STC explosively destructs outward due to the Maxwell stress and Joule heating.
Unfortunately, the STC does not work at its full potential because it has already expanded when the maximum magnetic field is reached.
Here, we propose an easy way to delay the expansion and increase the maximum field by using a mass-loaded STC.
By loading clay on the STC, the field profile drastically changes, and the maximum field increases by 4~\%.
This method offers an access to higher magnetic fields for physical property measurements.
\end{abstract}
\maketitle

\section{\label{Sec:1}Introduction}
The technical development of higher magnetic-field generation is always in demand for science and is still in progress.
After the pioneering development by Kapitza\cite{1924_Kap}, the field generation for a short time, known as the pulsed-field technique, is widely used to generate high magnetic fields.
Currently, 100-T class fields are available by using non-destructive pulsed magnets\cite{2002_Bac, 2010_Zhe, 2012_Jai, 2013_Zhe, 2018_Bat}.
For higher fields, destructive methods are utilized.
At the Institute for Solid State Physics (ISSP) in Kashiwa, 1000-T class fields can be generated by using the electromagnetic flux compression (EMFC) technique\cite{2018_Nak}, which is applied for condensed matter physics\cite{2011_Miy, 2020_Nak, 2020_Mat}.
This technique, however, requires laborious preparations because all the setup inside the magnet is destroyed after the explosive field generation.

\begin{figure}[t]
\centering
\includegraphics[width=0.85\linewidth]{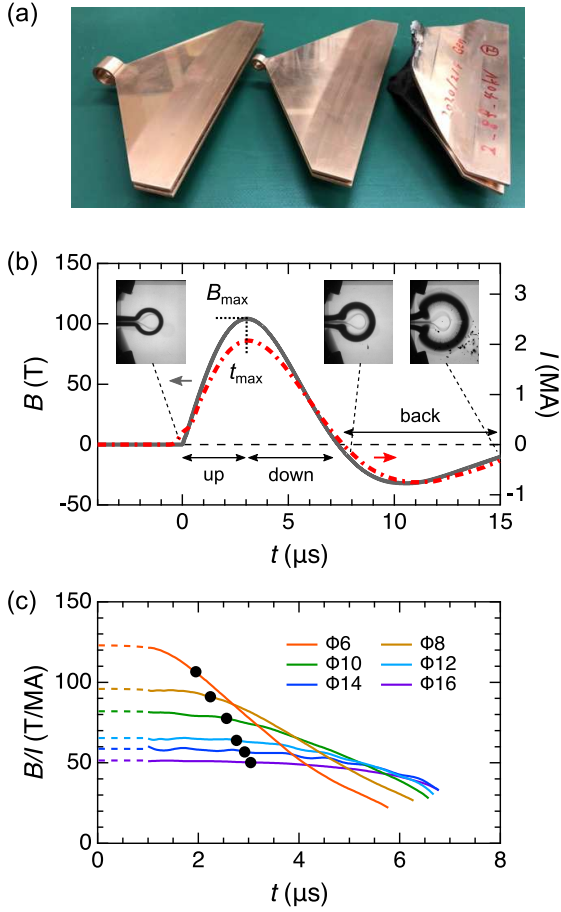}
\caption{(a) STCs used at ISSP. From the left, $\Phi$16, $\Phi$8, and $\Phi$8 STC after the field generation. (b) Typical waveforms of the magnetic field (gray line in the left axis) and current (red line in the right axis) as a function of time for the $\Phi$16 STC with 40~kV charging. The maximum field and the field-rising time are defined as $B_{\rm max}$ and $t_{\rm max}$, respectively. The flash-x-ray images of the $\Phi$16 STC during the field generation are also shown. (c) The field-current ratio $B/I(t)$ profiles for various sizes of STCs with 40~kV charging. The dashed lines are the guide for eyes. The black circles denote the $t_{\rm max}$ point for each size of STCs.} 
\label{Introduction}
\end{figure}

Single-turn coil (STC) technique is a convenient method to generate ultrahigh magnetic fields in a semi-destructive way.
The development of this technique dates back more than 60 years\cite{1957_Fur, 1969_She, 1973_Her, 1974_Her, 1997_Por, 1999_Por} and was initiated at ISSP in the 1980s\cite{1985_Nak, 1989_Miu, 1994_Miu, 2001_Miu, 2003_Miu}.
As shown in Fig.~\ref{Introduction}(a), the copper sheet with 3-mm thickness is bend to form the single-turn-coil shape.
By discharging $\sim$2~MA to the STC, magnetic fields above 100 T can be generated relatively easily.
The typical waveforms of the magnetic field ($B$) and the current ($I$) are displayed in Fig.~\ref{Introduction}(b).
The magnetic field reaches the maximum $B_{\rm max}$ in $t_{\rm max}=$ 2--3~$\mu$s and the (positive) field-duration time is 6--8~$\mu$s.
With a small-diameter STC, $t_{\rm max}$ becomes short and $B_{\rm max}$ exceeds 200~T.
During the field generation, the STC explosively destructs outward due to the Maxwell stress and Joule heating while keeping the sample and measurement setup inside the STC intact.
The deformation of the STC leads to a decrease in the field-current ratio $B/I$ already at $t_{\rm max}$, as shown in Fig.~\ref{Introduction}(c).

The STC system is designed on the basis of the following concepts.
First, the discharge needs to be fast enough compared to the coil destruction.
Otherwise, $B/I$ drops rapidly before $t_{\rm max}$, resulting in the reduction of $B_{\rm max}$.
To achieve the fast discharge, a low-inductance capacitor bank and the ``single''-turn coil (low inductance) are employed.
Second, the STC has to be made by a ``thin'' copper sheet to increase the current density near the sample space and to raise the field-generation efficiency, i.e. $B/I$.
However, if the copper sheet is too thin, the STC would deform too fast due to the insufficient mass.
As a compromise, the thickness of 3~mm has been adopted.

To increase $B_{\rm max}$ of the STC technique, it is important to keep a high $B/I$ ratio during the field generation.
In other words, the deformation of the coil needs to be suppressed.
From this viewpoint, the STC with larger inertia has an advantage to confine the Maxwell stress.
For example, $B_{\rm max}$ can be slightly increased with a STC made of higher density metal (e.g. tantalum) than copper \cite{1985_Nak}.
Another possibility is to drive a massive STC.
By employing a giant STC with 25-mm bore and the EMFC capacitor bank (40~kV, 5~MJ), the destruction of the coil is delayed up to $\sim$100~$\mu$s \cite{2010_Kin}.
However, the magnetic field reaches at most $B_{\rm max} \approx 100$~T because of the low $B/I$ ratio for such a large bore.

In this paper, we propose a simple way to increase $B_{\rm max}$ of the STC technique by using a mass-loading effect.
By coupling external mass to the STC, the inertia increases and the expansion of the coil can be delayed.
Judging from the $B/I(t)$ profiles shown in Fig.~\ref{Introduction}(c), we expect that there is room to raise $B_{\rm max}$ by a few percent especially for the small-diameter STCs.
We note that a similar idea has been already applied to the EMFC technique, where the outer coil made of steel is used as the external mass to suppress the destruction of the primary coil and efficiently concentrate the magnetic flux inside the liner \cite{2011_Tak}.

\section{\label{Sec:2}Methods}

\subsection{\label{Sec:2-1}Experimental detail}

The magnetic fields were generated with the charging voltage of 40.0~kV in the horizontal STC system at ISSP.
The capacitance of the system was 0.16~mF and the residual inductance was 16.5~nH.
The STCs with different diameters from $\Phi$6 to $\Phi$16 were employed to compare the mass-loading effect.
Here, the coil size is defined by the inner diameter, and the width of each coil is the same as the inner diameter.
The magnetic fields were measured by a calibrated pickup coil with the cross-sectional area of $\sim$1~mm$^{2}$, which was placed at the center of the STC.
The field profile $B(t)$ was obtained by integrating the pickup voltage $dB/dt(t)$ as a function of time $t$.
Hereafter, three regions of the $B(t)$ profile are termed as ``up'', ``down'', and ``back'' sweeps for the following discussions [Fig.~\ref{Introduction}(b)].
The current $I(t)$ flowing through the STC was measured by a Rogowskii coil located at the current collector plate of the horizontal STC system.

\begin{figure}[t]
\centering
\includegraphics[width=\linewidth]{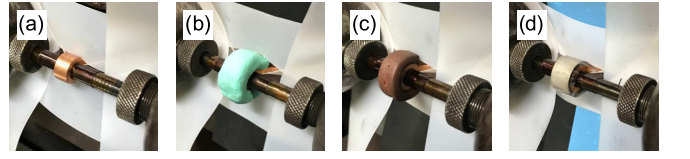}
\caption{Photos of the $\Phi$8 STC (a) without mass-loading, (b) with clay\#1, (c) with clay\#2, and (d) with a SUS ring.} 
\label{Photo}
\end{figure}

Two kinds of clay (\#1 and \#2) were employed as the insulating mass-loading materials.
Their properties are summarized in Table \ref{tab:table1}.
The sound velocity of each clay was estimated by the ultrasound measurement at the frequency of 100~kHz.
Clay\#1 mainly contains CaCO$_{3}$ and stone powder, whereas clay\#2 contains tungsten powder.
Note that the resistivity of clay\#2 is larger than 1~M$\Omega \cdot$cm.
Hence, both clays can be regarded as insulator.
The density of clay\#2 is more than three times as high as that of clay\#1, and comparable to that of copper.
Both clays can be easily deformed by hands and sticked on the surface of the STC as shown in Figs.~\ref{Photo}(b) and (c).
Additionally, stainless steel (SUS) was also employed as the metallic mass (only for the $\Phi$8 STC), whose results are briefly mentioned in Sec.~\ref{Sec:4-2}.
A SUS ring with the weight of 5.4~g was processed to fit the shape of the STC as shown in Fig.~\ref{Photo}(d).
Insulating Kapton tape with the thickness of 0.1~mm was inserted between the STC and the SUS ring.

\begin{table}[h]
\caption{Mass density ($\rho$) and sound velocity ($v$) of the mass-loading materials at ambient conditions. The densities of the stainless steel (SUS) and copper are also listed for comparison.}
\begin{tabular}{c | c c c}
\hline\hline
 & ~$\rho$ (g/cm$^3$)~ & ~$v$ (km/s)~ & Remarks \\
\hline
~Clay\#1~ & 1.9  & 1.4 & CaCO$_{3}$ and stone powder \\
~Clay\#2~ & 7.1 & 0.8 & Tungsten powder \\
~SUS~ & 7.8 & & Iron, Chromium, etc. \\
~Copper~ & 9.0 & & STC \\
\hline\hline
\end{tabular}
\label{tab:table1} 
\end{table}

\subsection{\label{Sec:2-2}Evaluation method of \mbox{\boldmath{$\Delta B_{\rm max}$}} and \mbox{\boldmath{$\Delta t_{\rm max}$}}}

Even when the magnetic fields are generated with the same design STC and charging voltage, there are slight variations in the measured $B_{\rm max}$.
This is attributed to several reasons, such as the STC-shape imperfection, position and calibration errors of the pickup coil, differences in the measurement environment (e.g. temperature), small fluctuations in the charging voltage, and so on.
Since the effect of the calibration error of the pickup coil is usually dominant among them, it should be ruled out in order to evaluate the mass-loading effect on $B_{\rm max}$ precisely.
Accordingly, we successively performed two identical shots without and with mass-loading using the same pickup coil, focusing on the relative change in the field strength, $\Delta B_{\rm max}$.
Due to slight differences in the performance of each STC, a variation of $B_{\rm max}$ of approximately 1~\% has been empirically observed for several shots with the same setup.

On the other hands, the observed $t_{\rm max}$ is not affected by the pickup coil calibration.
Hence, $\Delta t_{\rm max}$, the relative change of $t_{\rm max}$ with mass-loading, is evaluated by comparing with the statistical value of $t_{\rm max}$ without mass-loading, which is shown in Table~\ref{tab:table2} for each STC size.
Unfortunately, since not all the experiments were successful in measuring $B_{\rm max}$, some data sets could not be used for the $\Delta B_{\rm max}$ evaluation.
However, $\Delta t_{\rm max}$ can be evaluated for all the data with mass-loading, allowing us to collect many $\Delta t_{\rm max}$ data, as shown in Fig.~\ref{Experiment_plot}.
It is noteworthy that the $t_{\rm max}$ increase is indirect evidence, but supports the increase in $B_{\rm max}$.

\begin{table}[t]
\caption{Statistical average and error of the field-rising time $t_{\rm max}$ with the charging voltage of 40.0~kV for each size of the STC without mass-loading. All of the values are calculated from the sampling number of $N=5$ in the latest experiments. The errors are estimated by a 95~\% confidence interval assuming the $t$-distribution.}
\begin{tabular}{c | c c}
\hline\hline
$\Phi$ & $t_{\rm max}$ ($\mu$s) & ~~error (\%)~~ \\
\hline
~~16~~ & ~~$3.021 \pm 0.009$~~ & ~~0.29~~ \\
~~12~~ & ~~$2.753 \pm 0.024$~~ & ~~0.88~~ \\
~~8~~ & ~~$2.254 \pm 0.021$~~ & ~~0.93~~ \\
~~6~~ & ~~$1.940 \pm 0.025$~~ & ~~1.27~~ \\
\hline\hline
\end{tabular}
\label{tab:table2} 
\end{table}

\section{\label{Sec:3}Results}

\begin{figure}[t]
\centering
\includegraphics[width=\linewidth]{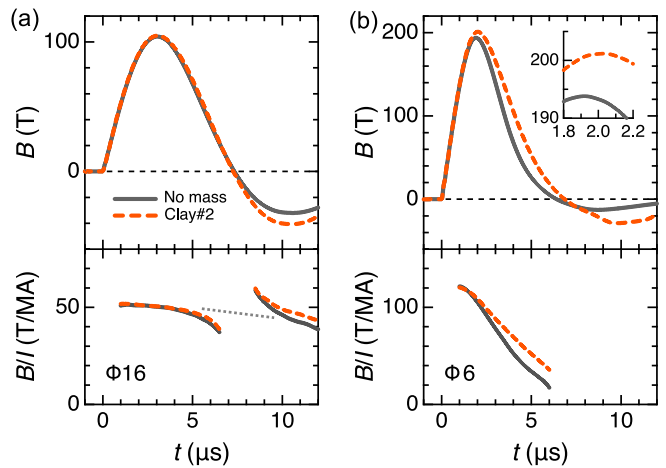}
\caption{(a) The magnetic-field waveforms (upper panel) and the $B/I(t)$ profiles (lower panel) generated by the $\Phi$16 STC without mass-loading (gray solid line) and with 15~g of clay\#2 (orange dashed line). (b) The magnetic-field waveforms (upper panel) and the $B/I(t)$ profiles (lower panel) generated by the $\Phi$6 STC without mass-loading (gray solid line) and with 60~g of clay\#2 (orange dashed line). The inset in (b) shows the enlarged view of $B(t)$ near $B_{\rm max}$ for the $\Phi$6 STC.}
\label{Experiment}
\end{figure}

First, let us focus on the mass-loading effect for the large-diameter STCs.
Figure~\ref{Experiment}(a) shows the typical $B(t)$ and $B/I(t)$ profiles obtained for the $\Phi$16 STC without (gray solid line) and with mass-loading (orange dashed line).
Here, 15~g of clay\#2 was employed.
Some parts of the $B/I(t)$ data are removed for the following reasons: (i) They were greatly affected by the initial noise before 1~$\mu$s. (ii) They diverge around 7~$\mu$s (The dotted line is drawn as a guide for eyes).
As shown in the $B(t)$ data, the magnetic-field waveform exhibits little change in the up and down sweeps by loading the mass.
The increase rates of $B_\mathrm{max}$ and $t_\mathrm{max}$ were only $\Delta B_\mathrm{max}=0.7$~\% and $\Delta t_\mathrm{max}=0.6$~\%, respectively.
This result indicates that the original mass of the standard $\Phi$16 STC is large enough to keep a high $B/I$ ratio until the maximum field is reached [Fig.~\ref{Introduction}(c)].
Notably, the mass-loading effect becomes clear in the back sweep, yielding the increase in the ``negative'' peak field by 25~\%, from $-32$~T to $-40$~T.
Such a significant increase in the negative peak field indicates that the field duration time can be lengthened by loading the external mass.

More pronounced mass-loading effect are observed for the small-diameter STCs.
Figure~\ref{Experiment}(b) shows the typical $B(t)$ and $B/I(t)$ profiles obtained for the $\Phi$6 STC without (gray solid line) and with mass-loading (orange dashed line).
Here, 60~g of clay\#2 was employed.
Note that the $B/I(t)$ data after 7~$\mu$s are removed because the $\Phi$6 STC is almost completely destroyed in this time region.
As shown in the $B(t)$ data, the two curves start to deviate around 1.5~$\mu$s and the difference is more pronounced in the down sweep.
The increase rates of $B_\mathrm{max}$ and $t_\mathrm{max}$ were $\Delta B_\mathrm{max}=3.7$~\% and $\Delta t_\mathrm{max}=4.6$~\%, respectively, which are much larger than the experimental errors.
Besides, the $B/I(t)$ profile with mass-loading exhibits a modest decrease compared to the normal case, indicating that the deformation of the STC is reduced by loading the external mass.

The mass-loading effect can be seen even after the field generation.
With the STC technique, the setup inside the coil usually survives after the explosive field generation because the Maxwell stress only expands the coil outward.
For the large-diameter STCs ($\Phi$16 and $\Phi$12) with mass-loading, the situation was the same.
However, for the small-diameter STCs ($\Phi$8 and $\Phi$6) with mass-loading, the setup inside the coil was heavily damaged after the field generation, indicating that the explosion occurs not only outward but inward.
The inward explosion can be understood by the law of action and reaction.
When the STC explodes, the volume of copper rapidly increases because of the Joule heating [see the images in Fig.~\ref{Introduction}(b)].
Hence, if the STC is surrounded by the external mass, the explosion is confined and the blast wave might go inward.

\begin{figure}[t]
\centering
\includegraphics[width=\linewidth]{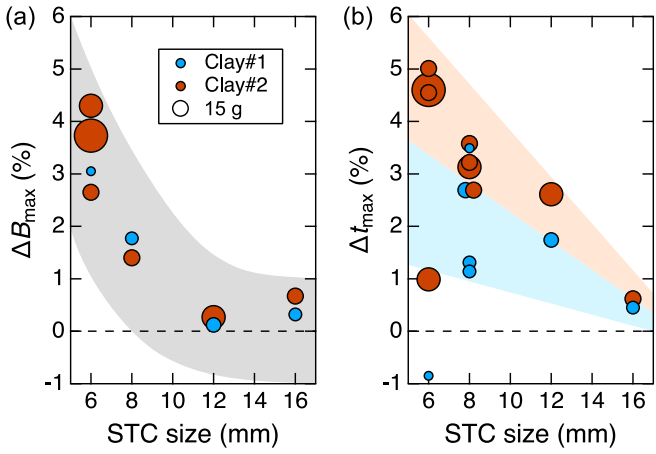}
\caption{Increase rates of (a) the maximum field ($\Delta B_\mathrm{max}$) and (b) the field-rising time ($\Delta t_\mathrm{max}$) by mass-loading STCs with different inner diameter. The area of the circle is proportional to the weight of the clay (see the legend for 15 g weight). The shadowed band indicates the trend of the plots with the expected error range. For the definitions of $\Delta B_\mathrm{max}$ and $\Delta t_\mathrm{max}$, see Sec.~\ref{Sec:2-2} for details.}
\label{Experiment_plot}
\end{figure}

Figure~\ref{Experiment_plot} summarizes the observed $\Delta B_\mathrm{max}$ and $\Delta t_\mathrm{max}$ with various mass-loading conditions.
Here, we plot all the experimental points for $\Delta t_\mathrm{max}$, but not for $\Delta B_\mathrm{max}$.
This is because $\Delta t_\mathrm{max}$ is not affected by the pickup coil calibration, as discuss in Sec.~\ref{Sec:2-2}.
The cyan and brown circles correspond to the results for clays\#1 and \#2, respectively, and the area of the circle indicates the amount of the clay.
The shadowed bands are the eye guide with the expected error range.
As shown in Fig.~\ref{Experiment_plot}(b), although some data exhibit a large variation possibly because of the slightly different STC shape, we can see the tendency that $\Delta t_\mathrm{max}$ monotonously increases as the STC size decreases.
This is reasonable because the deformation of the coil is faster for the small-diameter STCs and the external mass can delay the deformation more effectively.
In addition, $\Delta t_\mathrm{max}$ for clay\#2 is almost twice as large as for clay\#1.
This would reflect the difference in the density of the clay.
In contrast, $\Delta t_\mathrm{max}$ seems to be independent of the amount of the clay, suggesting that only a tiny layer of the clay effectively couples to the STC even though a larger amount of clay is attached.
The effectively-coupled amount of clay is estimated in Sec~\ref{Sec:4-1}.
Despite of the distinct mass-loading effect on $t_\mathrm{max}$, the increase in $B_\mathrm{max}$ is not clear for the large-diameter STCs ($\Phi$16 and $\Phi$12) as shown in Fig.~\ref{Experiment_plot}(a).
Nevertheless, the sufficient $\Delta B_\mathrm{max}$ increase was well reproduced for the small-diameter STCs ($\Phi$8 and $\Phi$6), confirming our expectation that $B_\mathrm{max}$ can be increased using this method.

\section{\label{Sec:4}Discussions}

\subsection{\label{Sec:4-1}Effectively-coupled mass}

\begin{figure}[t]
\centering
\includegraphics[width=\linewidth]{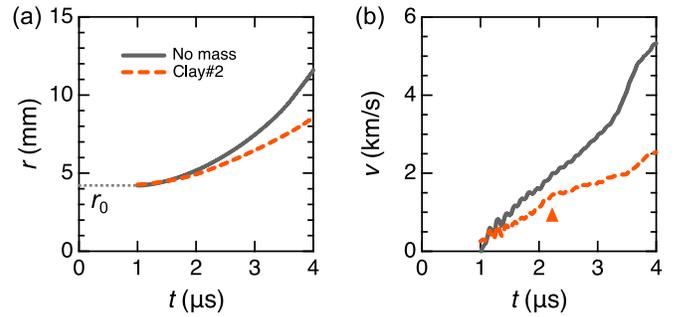}
\caption{Time dependences of (a) the coil radius and (b) the velocity of the coil expansion for the $\Phi$6 STC without mass-loading and with 60~g of clay\#2, which are estimated from the experimental results shown in Fig.~\ref{Experiment}(b). See text for details.}
\label{Effective_mass}
\end{figure}

From the experimental $B/I(t)$ profiles, we can extract information on the radius of the STC, $r(t)$, during the field generation.
Here, we focus on the cases of the $\Phi$6 STC without mass-loading and with 60 g of clay\#2.
As a simplified model for the STC, we consider a cylinder-shaped coil with the initial radius of $r_{0}$, the width of $w$, and no thickness.
Based on this model, the coil radius is obtained by the Ampere's law as $B/I(t)={\mu_{0}}/[2{\sqrt{r(t)^{2}+(w/2)^{2}}}]$.
Now, the initial coil radius $r_{0}$ is estimated from the experimental value of $B/I(0) \approx 123$ T/MA as $r_{0} \approx 4.1$~mm, which reasonably coincides with the real STC radius (the inner radius of 3 mm and the outer radius of 6 mm).
If we fix the coil width as $w=6$~mm during the field generation, $r(t)$ is obtained as Fig.~\ref{Effective_mass}(a).
By taking the time derivative of $r(t)$, the speed of the coil expansion along the radial direction, $v(t)$, is obtained as Fig.~\ref{Effective_mass}(b).
Note that the calculated $v(t)$ would be overestimated in the time range of $t \gg 3$~$\mu$s.
It is because the actual deformation of the STC occurs also for the axial direction and the simplified model adopted here would not be a good approximation.
Nevertheless, this analysis is considered to be sufficient to compare the ways of the STC expansion with and without mass-loading around $t_{\rm max} \approx 2$~$\mu$s.

Here, let us quantitatively discuss the mass-loading effect.
As shown in Fig.~\ref{Effective_mass}(b), the $v(t)$ profile without mass-loading exhibits an almost linear increase in the time range of $1 \sim 3$~$\mu$s.
On the other hand, as denoted by a triangle in Fig.~\ref{Effective_mass}(b), the $v(t)$ profile with clay\#2 exhibits a hump-like behavior around $t_{\rm h}=2.2$~$\mu$s.
This feature is accountable by considering the dynamical deformation of the clay.
Before $t_{\rm h}$, the slope of $v(t)$ with clay\#2 is roughly two-thirds compared with that without mass-loading, indicating that the effective coil-mass becomes approximately 1.5 times larger by loading clay\#2 in this time range.
Considering that the weight of the cylinder part of the $\Phi$6 STC is approximately 4~g, the effectively-coupled amount of clay\#2 is estimated to approximately 2~g, which is much less than the attached amount of 60~g.
This is because the stress is confined in the inner layer of the clay due to its high compressibility, i.e. the slow sound velocity (0.8~km/s at ambient conditions).
However, after $t_{\rm h}$, the slope of $v(t)$ with clay\#2 is more suppressed, suggesting the increase of the effectively-coupled mass.
This may be due to the formation of the shock wave\cite{2017_And}, which is produced when the speed of the STC expansion $v(t)$ becomes comparable to the sound velocity of clay\#2.
In such a situation, the stress would concentrate on the shock front.
Since $v(t)$ increases even after $t_{\rm h}$, the shock front is expected to develop, resulting in a further coupling of the external mass.
As mentioned above, it is difficult to quantitatively estimate the effectively-coupled mass in the time range of $t \gg 3$~$\mu$s.
Experimentally, it was found that the $B/I(t)$ profiles for the $\Phi$6 STC with 15~g and 30~g of clay\#2 almost overlap with that with 60~g of clay\#2 until 6~$\mu$s (not shown).
Therefore, we infer that less than 15~g of clay\#2 can finally coupled to the $\Phi$6 STC during the field generation. 

\subsection{\label{Sec:4-2}Proposal of mass-loading materials}

To improve the mass-loading effect, our results suggest to use materials with the following properties.
First, materials with a high mass density are favored.
Our experimental results for clays\#1 and \#2 (Fig.~\ref{Experiment_plot}) clearly indicate that the higher mass density leads to higher field generation.
Second, materials with a faster sound velocity (lower compressibility) are favored.
Our analysis revealed that only a tiny layer of the clay couples as the external mass during the field generation.
If the sound velocity of the material is faster, the thickness of the effectively-coupled mass is expected to increase.

Accordingly, metallic materials (such as steel) can be good candidates to satisfy these requirements.
As already introduced in Sec.~\ref{Sec:2-1}, the experiment with stainless steel (SUS) as the external mass was also performed.
However, the result shows the reduction of $B_{\rm max}$ with a significant change of the field waveform (not shown).
This is probably due to the electrical insulation breakdown between the two metallic parts, the STC and the SUS ring.
When the electric current passes through the SUS ring, the field generation efficiency $B/I$ decreases, because the SUS ring locates outside the STC.
With the huge Maxwell stress and high voltage, keeping the electrical insulation between the STC and SUS ring was impossible.
Thus, metals are not realistic materials as the external mass for our purpose.
As the insulating external mass, ceramic materials might also be the candidates.
However, the densities of common ceramics (SiO$_2$, Al$_2$O$_3$, and so on) are not high as metals and clay\#2.
Besides, processing the ceramics to fit the STC is rather challenging and unrealistic.

From the above discussions, we conclude that clay\#2 is a good compromise to our requirements, the high density and the insulating properties.
With the proposed way, the destruction of the STC is reasonably delayed and the higher field generation is realized.
Besides, we emphasize that the concept of the mass-loaded STC, which only changes the effective mass but does not affect the electrical circuit, is relatively simple and safe.
Our strategy would be useful for the experiments where a bit higher-field generation is required in the STC system.

\section{\label{Sec:5}Conclusion}
We experimentally demonstrated that the field profile in the STC system can be modified by loading clay as the external mass.
For the $\Phi$6 STC, 4~\% and 5~\% increase in $B_{\rm max}$ and $t_{\rm max}$ were achieved, respectively.
Although a STC with smaller diameter than 6~mm can generate 300-T class magnetic fields without the present method \cite{1999_Por}, such a STC is not useful for the physical property measurements because of the limited sample space and the short measurable time.
Importantly, the present work shows that $B_{\rm max}$ as well as $t_{\rm max}$ can be increased for $\Phi$8 and $\Phi$6 STCs.
At ISSP, the physical property measurements using the STC technique has been so far limited up to $\sim$240~T (using the $\Phi$6 STC with the charging voltage of 50~kV) \cite{2020_Gen}.
The proposed strategy using the mass-loading effect expands the potential of the STC technique and opens the door for higher-field science.

\section*{Acknowledgments}
This work was partly supported by the JSPS KAKENHI Grants-In-Aid for Scientific Research (No. 18H01163, No. 19K23421, No. 20J10988, No. 20K14403, and 20K20892).
M.G. was supported by the JSPS through a Grant-in-Aid for JSPS Fellows.

\section*{Data Availability Statement}
The data that support the findings of this study are available from the corresponding author upon reasonable request.

\end{document}